# Electrostatic Imaging of Encapsulated Graphene

Michael A. Altvater[1], Shuang Wu[1], Zhenyuan Zhang[1], Tianhui Zhu[1i], Guohong Li[1],
Kenji Watanabe[2], Takashi Taniguchi[2] and Eva Y. Andrei[1]*

[1] Department of Physics and Astronomy, Rutgers, the State University of New Jersey, 136 Frelinghuysen Rd, Piscataway, New Jersey 08854, USA
[2] National Institute for Material Science, 305-0044 1-1 Namiki Tsukuba, Ibaraki, Japan

E-mail: eandrei@physics.rutgers.edu

**Abstract**

Devices made from two-dimensional (2D) materials such as graphene and transition metal dichalcogenides exhibit remarkable electronic properties of interest to many subdisciplines of nanoscience.  Owing to their 2D nature, their quality is highly susceptible to contamination and degradation when exposed to the ambient environment. Protecting the 2D layers by encapsulation between hexagonal boron nitride (hBN) layers significantly improves their quality.  Locating these samples within the encapsulant and assessing their integrity prior to further processing then becomes challenging.  Here we show that conductive scanning probe techniques such as electrostatic force and Kelvin force microscopy makes it possible to visualize the encapsulated layers, their charge environment and local defects including cracks and bubbles on the sub-micrometer scale.  Our techniques are employed without requiring electrical contact to the embedded layer, providing valuable feedback on the device's local electronic quality prior to any device etching or electrode deposition.  We show that these measurement modes, which are simple extensions of atomic force microscopy, are perfectly suited for imaging encapsulated conductors and their local charge environments.

## 1. Introduction

2D materials can display remarkable electronic properties, but with all the atoms at the surface these properties are easily obscured by scattering off substrate-induced random potentials or adsorbed species.  It is therefore desirable to protect these layers by using flat, inert substrates or by removing the substrate altogether (1-10).  Hexagonal boron nitride (hBN) is considered an ideal substrate because of its atomic flatness and its ability to segregate hydrocarbon contaminants into bubbles.  Outside these bubbles, hBN provides a pristine surface with significantly reduced charge fluctuations, thus providing access to the low energy electronic properties of graphene and other 2D layers placed on top.  Development of methods to pick up and transfer 2D-materials using hBN has led to ultra-clean, high-quality, encapsulated devices such as FETs (11), photodetectors (12), and light emitters (13).  Graphene encapsulated in hBN (14) has provided access to a wide range of intrinsic properties including micron-scale ballistic transport (15), electron optics (16, 17), magnetic focusing (18), and Moiré superlattices which exhibit interesting magneto-transport properties (19-21). Adding a second graphene layer on top of the first further allows one to tune the band structure by controlling the relative angle between the graphene flakes (22, 23).  Devices made from these structures are quickly gaining attention in the physics research community after the observation of correlated electronic ground states at low temperatures (24, 25).

A common challenge in the fabrication of any 2D heterostructure is evaluating and maintaining the cleanliness of the planar interfaces.  Protecting the electronic properties of a 2D sample by encapsulation is a delicate process.  The quality of such encapsulated devices depends on the location and number of contaminants trapped within the device and on the reliability of the electrical contacts.   Typical 2D heterostructures can be constructed in air or in an inert atmosphere glovebox (for reduced contamination) using the polymer stamp method (described in Supporting Information) (15, 26).

We observe that even with substrate components annealed and encapsulation performed in a glovebox, samples often exhibit collections of hydrocarbons in bubbles within the structure, electrostatic charging, adsorbates, strain build-up, and even formation of cracks and tears during encapsulation. Typically, optical microscopy and AFM topography are used to determine regions of the device which are free of bubbles and wrinkles before these relatively defect free regions are isolated by plasma etching (see Figure 1a); to produce a high-quality device.

Previous works have determined the degree of electronic disorder by measuring the spatial variation of the Fermi level of graphene placed on top of hBN and contacted with an electrode by STS mapping (3, 4, 6, 8, 10). Transport measurements probe the degree of low-energy electronic scattering by measuring the energy spread of the high resistive state at charge neutrality in encapsulated graphene layers after employing plasma etching and electrode depostition (27). Here we utilize non-contact AFM-based electrostatic measurements of potential variations near the surface of encapsulated graphene heterostructures. Using simple analysis, we relate these measured quantities to the charge inhomogeneity in the buried graphene layer. Further, we employ high-resolution electrostatic imaging to locate the gr. Our measurements technique benefits from not requiring electrical contact to the graphene layer, making it possible to image and characterize other encapsulated samples non-invasively during different stages of device fabrication.

## 2. Methods

Devices are fabricated with the standard polymer stamp method using the strong adhesion between graphene and hBN (see Supporting Information) and placed onto p-doped silicon wafers with a capping layer of 300nm $SiO_2$. Devices were constructed using remotely controlled micromanipulator stages inside an argon-atmosphere glovebox with oxygen and water levels maintained below 0.1ppm.

Samples are measured using an NT-MDT SolverNEXT SPM system with gold coated AFM tips with typical cantilever stiffness of 3.5N/m (HA-NC probes from K-Tek nano). Relative humidity in the AFM measurement chamber is maintained below 5% by a slow flow of dry nitrogen gas in order to avoid effects of water accumulation on the sample surface.

Electrostatic measurements are performed in the two-pass scheme where in the first pass, the topography is measured using non-contact AFM; on the second pass, electrostatic forces are monitored with the tip retracted 10nm from the measured sample surface. We note that it is important to ground the doped silicon backgate during measurement to allow quantitative analysis and to avoid charging/discharging of the tip or sample.

High-resolution electrostatic imaging is performed by measuring the electrostatic force gradient near the surface using electric force microscopy (EFM). The cantilever is excited using a piezo block (as in non-contact topography mode) and a DC bias is applied. The applied bias modifies the electrostatic force felt by the AFM tip and induces a phase shift between the driving signal and the cantilever oscillation which can be related to the effective tip-backgate capacitance and the surface potential by

$$\Delta\phi = -arcsin\left(\frac{Q}{k}\frac{\partial F_{es}(V_b)}{\partial z}\right) \quad (1)$$

where $Q$ and $k$ are the quality factor and spring constant of the AFM cantilever, respectively, and $F_{es}(V_b)$ is the electrostatic force experienced by the tip due to the sample:

$$F_{es}(V_b) = -\frac{\partial C}{\partial z}(V_b - V_S)^2. \quad (2)$$

Here, C is the total capacitance between the tip and backgate, $V_b$ is the applied tip bias, and $V_S$ is the potential at the sample surface.

The advantages of imaging with EFM are two-fold. Firstly, the measurement signal is nearly quadratic in the applied bias. This allows one to controllably tune the contrast between regions with different effective capacitances. Second, the phase shift measured is proportional to the second derivative of the capacitance which is highly localized to the tip apex; providing superior resolution to measurement modes which probe the force rather than the force gradient (28-31).

Surface potential measurements are performed with amplitude modulated Kelvin force probe microscopy (AM-KPFM) where DC and AC signal is applied to the tip at frequency $\omega$, inducing frequency dependent components to the electrostatic force. The force component proportional to $\cos\omega t$ is given by

$$F_\omega(z) = -\frac{\partial C}{\partial z}(V_{DC} - V_S) \cdot V_{AC}\cos\omega t, \quad (3)$$

where $V_{DC}$ and $V_{AC}$ are the applied DC and AC components of the bias applied to the tip. During the KPFM measurement, this component is monitored and the DC bias is controlled by feedback to nullify $F_\omega(z)$, ie. find $V_{DC}$ such that $V_{DC} = V_S$, determining the surface potential of the sample.

On Si/SiO2, the sample surface potential can be related to work function difference between tip and backgate modified by charges within the structure:

$$V_S = \frac{\Delta\Phi}{e} - \frac{q}{C_b}, \quad (4)$$

where $\Delta\Phi$ is the workfunction difference between the gold-coated tip and the silicon backgate, e is the charge of an electron, q is ths total charge trapped within the structure beneath the tip, and $C_b$ is the capacitance between the localized charge and the grounded backgate (32).

Now, assuming graphene forms parallel plate capacitor with the backgate and that the charge in graphene is uniformly distributed (i.e. tip-graphene image interactions are negligible), we find that the difference in surface potential measured between regions with encapsulated graphene present and regions without graphene can be given simply by

$$\Delta V_S = -\frac{\varepsilon_{hBN}d_{ox} + \varepsilon_{ox}t_{bot}}{\varepsilon_{ox}\varepsilon_{hBN}\varepsilon_0}en_{2D}, \quad (5)$$

where $d_{ox}$ is the thickness of the silicon oxide (300nm), $\varepsilon_{ox}$ is the dielectric constant of silicon oxide (about 3.9), $\varepsilon_{hBN}$ is the out-of-plane dielectric constant of hBN (about 3.8), $\varepsilon_0$ is the permittivity of free space, $t_{bot}$ is the thickness of the hBN flake beneath the graphene, and $n_{2D}$ is the 2D charge density of the buried graphene layer. Thus, after measuring the thickness of the bottom hBN flake by AFM topography we are able to extract the 2D carrier concentration of the encapsulated graphene layer by simply comparing its local surface potential to that of the surrounding regions.

## 3. Results and Discussion

### 2.1 Electrostatic Force Microscopy (EFM)

We first image the samples by EFM phase shift mapping. The optical micrograph of an encapsulated double layer graphene sample is shown in Figure 1a. It consists of two stacked graphene monolayers (~zero-degree twist angle) and is left electrically isolated during this measurement. The first pass of the measurement records the sample topography and is displayed in Figure 1c,d. The encapsulated graphene flake is burried between a 71nm hBN flake underneath and a 50nm hBN flake on top. By applying a bias between the AFM tip and the silicon backgate, we are able to selectively image the buried graphene layer. In Figure 1e, we measure the EFM phase shift at a DC bias $V_b = -6V$ in the same region shown in Fig. 1c and observe a large phase shift difference when the tip passes above the region of encapsulated graphene.

Note that we are also able to observe the local gold backgate which runs beneath the sample in the phase shift measurement, showing that this method may be employed for other thin conducting materials.

We measure the phase shift vs bias in different regions of the device (Figure 1f): a region of bare hBN (no graphene), an encapsulated monolayer graphene flake, and one region where two graphene layers overlap (BLG). In each region, the signal minimum is related to the surface potential while the curvature relates to the second derivative of the local capacitance. The phase shift minima nearly coincide in all regions of the device; however, there is a large enhancement of $\frac{\partial^2 C}{\partial z^2}$ above the graphene layer compared to the surrounding

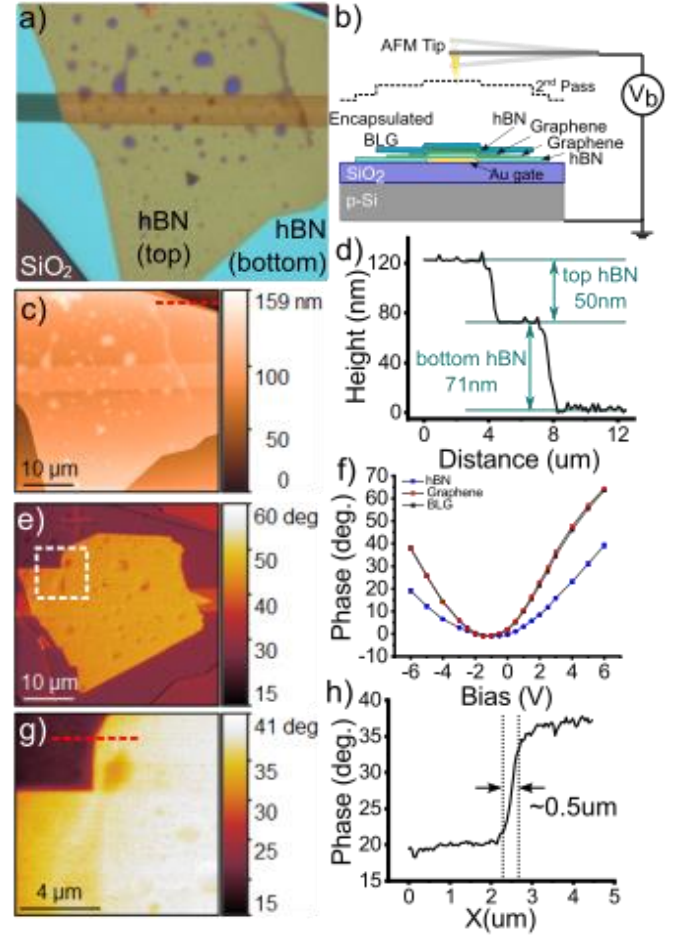

*Figure 1. Electrostatic Force Microscopy a) An optical microscope image of an encapsulated graphene device showing the exposed silicon dioxide substrate in the bottom left corner, the bottom hBN flake beneath the graphene layer appears light blue, and the hBN flake on top of graphene appears yellow. Additionally, bubbles which form within the structure during the encapsulation process can be seen as blue and brown spots and a thin gold backgate (15nm thick) runs beneath the sample and can be seen optically as an orange horizontal stripe. b) A schematic diagram of the measurement of an encapsulated bilayer graphene sample during the 2nd pass of the measurment. The tip is maintained at a distance above the sample surface, following the surface morphology, while a bias is used to probe the electrostatic properties of the structure. c) AFM topography of the device imaged in (a) d) A linecut through the topography image along the dashed line in (c) gives the thicknesses of the encapsulating hBN layers e)An EFM phase shift map in the same region as (c) displays a clear phase shift in the vicinity of the encapsulated graphene region f) The dependence of the measured phase shift on applied tip bias in different regions of the device shown in (e). The phase shift minimum relates to the surface potential and the curvature relates to the second derivative of capacitance through equations (1) and (2) g) A zoomed in EFM phase shift map at the edge of the encapsulated graphene region h) A*

*linecut through the EFM phase shift map along the dashed line in (g) shows the sharp edge resolution obtainable with this technique.*

regions both due to the screening ability of the graphene layer and due to the close proximity of the graphene to the tip. Thus, the contrast between the regions of encapsulated graphene and the surrounding hBN stack grows with increasing magnitude of $V_b$.

A zoomed-in phase shift map is taken at the corner of the device where the two encapsulated graphene monolayers overlap (figure 1g) and a linecut across the edge of the buried graphene is plotted in figure 1h. The EFM phase shift signal decays quickly away from the graphene edge (within 500nm), defining its position with high accuracy.

To get a feel for the effect of encapsulant thickness on the measurement signal, we make use of an encapsulated graphene sample which has a folded sheet of hBN on top. The optical image and AFM topography is shown in Figure 2a and b, respectively. The top hBN flake has folded on top of itself twice providing three separate regions with different encapsulating thicknesses determined from a linecut through the AFM topography (teal line in Fig. 2b) shown in Figure 2d. Three such regions are marked in the EFM phase shift map (teal stars in Fig. 2c) and their measured phase shift is plotted vs encapsulant thickness in Figure 2e and shows a monotonic decrease of the phase signal with hBN thickness. Similar to increasing the hBN thickness, we change the tip-sample separation (lift height) and monitor the phase shift above the encapsulated graphene and above bare hBN with a tip bias of $V_b = -3V$. We find that the contrast between encapsulated graphene and bare hBN decreases sharply for tip-sample distances between 10nm and few hundred nanometers, becoming immeasurably small for tip-sample distances larger than the thickness of the silicon oxide substrate.

*2.2 Kelvin Probe Force Microscopy (KPFM)*

Next we employ AM-KPFM to map the surface potential of the sample shown in Figure 1. The results are displayed in Figure 3. The region where graphene is encapsulated shows a lower surface potential with respect to the tip than the surrounding boron nitride idicating a difference in local charge density given by equation (5). By determining the thickness of the bottom hBN flake (from Fig. 1d) and evaluating using equation (5), we can determine the 2D charge density within the graphene layer to be $n_g = 0.92 \times 10^{10}\ 1/cm^2$. This level of doping is typical for encapsulated graphene devices fabricated in a dry atmosphere. Despite fabrication in a controlled environment using fresh, clean materials, we always observe some small amount of charge bound to the graphene layer. This charge density is large enough to provide the contrast observed in the surface potential map.

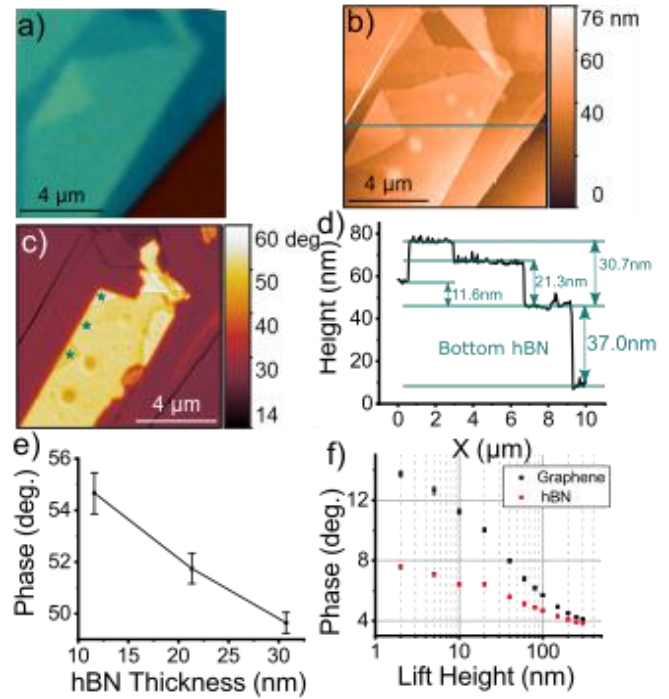

*Figure 2. Tip Height and hBN Thickness Dependence a) An optical image of an encapsulated graphene sample with a folded hBN flake on top. b) An AFM topography scan taken at the same location shown in (a) showing the structure of the folded hBN. c) The EFM phase shift map taken at $V_b$=-6V in the region shown in (a) and (b) shows an encapsulated graphene layer beneath the folded hBN flake. Regions of the graphene covered with different thicknesses of hBN are marked with stars. d) A linecut through the AFM topography scan along the teal line in (b). The graphene sits on top of a hBN flake 37nm thick with three different thicknesses of hBN on top: 11.6nm, 21.3nm, and 30.7nm. e) The cantilever phase shift recorded at $V_b$=-6V for three different thicknesses of encapsulating hBN marked in (c) shows a monotonic decrease of the force gradient as thickness increases. f) The cantilever phase shift vs. lift height taken at $V_b$=-3V above another encapsulated graphene sample (not shown) displaying the sharp decrease in contrast between the encapsulated graphene and surrounding hBN as the tip-sample separation grows large.*

Due to the charge trapped in the embedded graphene, we observe a surface potential contrast between the encapsulated monolayer and bilayer graphene. The difference in doping dependence of monolayer and bilayer graphene stems from the difference in low energy band structures. To maintain the same Fermi level, the bilayer graphene requires a larger charge density, as seen in Figure 3b. With the surrounding boron nitride stack as background, we find a charge carrier concentration in the bilayer region of $n_{BLG} = 1.16 \times 10^{10}\ 1/cm^2$. Note that from AFM topography (Figure 1c), neither the graphene edge nor the monolayer/bilayer boundary are visible, however, using KPFM we are able to select device regions to produce monolayer, bilayer, or a device across the boundary.

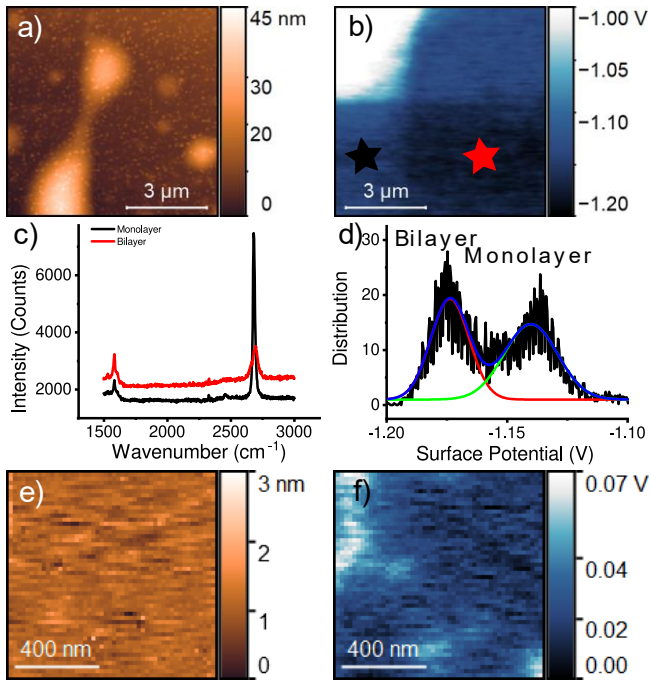

*Figure 3. Kelvin Probe Force Microscopy a) A surface potential map of the sample shown in Fig.1. The charge of the buried graphene layer locally modifies the measured surface potential. b) A zoom in to a small region outlined in (a) where two encapsulated graphene layers overlap. The band structure differences between monolayer and bilayer graphene leads to an observable difference in observed charge density when the overall graphene layer is charged. c) Raman spectrum taken at the location of the stars in (b) confirms that the contrast that we observe is due to the boundary between monolayer and bilayer graphene. Note that the graphene D-peak is not shown due to convolution with the hBN G-peak. d) A histogram of surface potential values measured in the monolayer and bilayer regions gives a measure of surface potential fluctuations. e) An AFM topography scan of an example region where there are few topographical features. f) The surface potential map in the same area as (e) shows large potential fluctuations which are not clearly associated with any topographical features. Scale is offset to reflect the magnitude of the surface potential fluctuations in this region.*

Upon fabricating and attaching electrodes to the graphene layer, the trapped charge will change due to the work function differences between the graphene layer and the contacting metal. Thus, the absolute charge density is not a good metric for device quality. Instead, we may look at the surface potential fluctuations within the encapsulated graphene region. The potential within the graphene layer is expected to be uniform and thus, any surface potential fluctuations must originate from unscreened charges within the structure. Surprisingly, despite the prescence of bubbles observed in sample topography (Fig. 3a), the measured surface potential seems locally unaffected. However, we do observe fluctations in the surface potential across the device surface refelcted in the RMS spread of the surface potential distribution (Fig. 3d). Additionally, in some regions of the device, we observe local potential fluctuations (up to 70mV) which are not observed to be associated with any topographical features (Fig. 3e,f). These features provide strong electronic scattering centers which might accidentally be included in a processed device if using AFM topography alone. It is well-known that charge traps, which are typically found within the insulating SiO2 substrate, the hBN encapsulant and at the various interfaces, provide a random potential and scattering centers which reduce the device's transport properties. We measure the RMS variation in surface potential in several regions of the device. For the monolayer (g) and bilayer (BLG) regions, we find RMS variations of $\delta V_g = 26.09 mV$ and $\delta V_{BLG} = 18.73 mV$ corresponding to charge variations of $\delta n_g = 1.51 \times 10^9 \; 1/cm^2$ and $\delta n_{BLG} = 1.08 \times 10^9 \; 1/cm^2$, respectively. The values obtained are comparable to those measured by transport in similar devices (27). Thus, we can use this metric to identify pristine regions of the sample that are suitable for further processing so as to maximize device quality.

We provide, in the Supporting Information, a diagrammitc description of the encapsulation procedure, measurements of dirty and defective samples identified by EFM, and finally, we demonstrate preliminary results regarding the effects of annealing on sample uniformity.

## 3. Conclusions

In conclusion, we have demonstrated that conductive scanning probe techniques are well suited for achieving high-contrast and high-resolution imaging of 2D conductive layers encapsulated in hBN. We show that the small interaction volume between the tip and sample enables the user to visualize edges of the embedded flake at the sub-micron scale. Employing surface potential mapping with KPFM, we identify the positions of charged contaminants within and on top of the heterostructure through fluctuations of the measured surface potential identifying electronically defective and pristine regions of the buried graphene. The techniques used are extensions of atomic force microscopy which are commercially available or can be implemented through simple upgrades to existing AFM systems. We expect that these methods will be valuable tools for fabricating high quality transport devices by avoiding contaminated and defective regions of the encapsulated flake(s). Finally, we hope that these techniques will lead to the production of pristine heterostructures, open new avenues for 2D device characterization methods, and provide opportunities for novel research.

## Acknowledgements

We gratefully acknowledge support from the Department of Energy (DOE-FG02-99ER45742 (E.Y.A.)) and the National Science Foundation (NSF EFRI 1433307 (M.A.A., T.Z.), NSF DMR 1708158 (G.L.), C/S NSF DMR 1337871(G.L.)). Authors would like to give special thanks to M. Surtchev of NT-MDT for his continued assistance establishing and troubleshooting our measurement system.  M.A.A. would also like to thank J. Duan for helpful discussions and motivation.

---

[i] Address listed location of where the work was performed. Current address is 351 McCormick Road, Charlottesville, VA 22904